\documentclass[12pt]{article}
\usepackage[utf8]{inputenc}
\usepackage[parfill]{parskip}
\usepackage{hyperref}
\usepackage{booktabs}
\usepackage{amsmath}
\usepackage{amsthm}
\usepackage{amsfonts} 
\hypersetup{colorlinks=true, urlcolor=blue}
\usepackage[margin=0.9in]{geometry}
\usepackage{enumitem}
\usepackage{mdframed}
\usepackage{graphicx}
\usepackage{amssymb}

\setlist[enumerate,1]{label = (\alph*)}
\setlist[enumerate,2]{label = \roman*.}

\newcommand{\R}{\mathbb{R}}

\newcommand{\supp}{\text{supp}}

\newtheorem{theorem}{Theorem}
\newtheorem{definition}{Definition}

\newtheorem{conjecture}{Conjecture}

\title{On the Exponential Circuit Imbalance of the Ben-Tal Nemirovski Approximation\thanks{Supported in part by a Discovery Grant and an Undergraduate Student Research Award from the Natural Sciences and Engineering Research Council (NSERC) of Canada.}}

\author{
    Jonah Bondar\thanks{Department of Combinatorics \& Optimization, University of Waterloo, 200 University Ave.\ W., Waterloo, ON, N2L 3G1, Canada, \texttt{jbondar@uwaterloo.ca}.}
    \and
    Stephen A. Vavasis\thanks{Department of Combinatorics \& Optimization, University of Waterloo, 200 University Ave.\ W., Waterloo, ON, N2L 3G1, Canada, \texttt{vavasis@uwaterloo.ca}.}
}
\date{\today}

\begin{document}

\maketitle

\begin{abstract}
Dadush et al.\ (2024) recently developed a scaling-invariant layered least squares algorithm for linear programming whose complexity depends on the optimal condition measure $\bar{\chi}_A^*$. Their work builds on Vavasis and Ye's (1996) algorithm whose running time depends only on the constraint matrix $A$ through the condition number $\bar{\chi}_A$. Monteiro-Tsuchiya (2003) defined the optimal condition number $\bar{\chi}_A^*$ as the maximum $\bar{\chi}_{AD}$ achievable over all positive diagonal column rescalings $D$. Dadush et al.\ (2024) introduced the optimal circuit imbalance measure $\kappa_W^*$, which serves as a lower bound for $\bar{\chi}^*_A$.

Instances with artificially large optimal circuit imbalance measures $\kappa_W^*$ can be easily constructed; however, finding naturally occurring examples where this optimal scaling-invariant measure grows exponentially is of independent interest. In this paper, we show that the Ben-Tal Nemirovski (BN) linear programming approximation of the unit disk provides such an example. By explicitly constructing circuits in the kernel of the BN formulation, we prove that the optimal circuit imbalance measure $\kappa_W^*$ grows exponentially in the number of approximation steps. Since $\kappa_W^*$ lower bounds $\bar{\chi}_A^*$, our result demonstrates that the BN approximation yields an exponentially ill-conditioned family of constraint matrices.
\end{abstract}

\section{Introduction}

Consider a linear program (LP) in standard equality form (SEF), defined as
\[
\min \, \{ c^\top x \mid Ax = b, \, x \ge 0 \},
\]
where $A \in \R^{m \times n}$ is the constraint matrix. 

The LP algorithm of Vavasis and Ye \cite{vavasis1996} achieves a running time depending only on the constraint matrix $A$, parameterized by the condition measure $\bar{\chi}_A$. Recently, Dadush et al.\ \cite{dadush2024} developed a scaling-invariant layered least squares (LLS) algorithm whose complexity depends instead on the optimally scaled measure, as defined by Monteiro-Tsuchiya \cite{doi:10.1137/S1052623401388926}
\[
\bar{\chi}_A^* = \inf_{D \in \mathbf{D}} \bar{\chi}_{AD},
\]
where $\mathbf{D}$ denotes the set of $n \times n$ strictly positive diagonal matrices. The Dadush et al.\ algorithm requires $O(n^{2.5} \log(n)\log(\bar{\chi}_A^* + n))$ iterations, and its analysis relies on bounding $\bar{\chi}_A^*$ via optimal circuit imbalances.

While one can artificially construct constraint matrices with exponentially large $\bar{\chi}_A^*$ (such as the tropical geometry example noted in Dadush et al.\ \cite[p.~143]{dadush2024}), identifying natural sources of such ill-conditioning is of independent interest. In this paper, we analyze the Ben-Tal Nemirovski (BN) approximation of the unit disk \cite{bental2001}, a standard polyhedral approximation for second-order cone constraints. We prove that the BN approximation has an optimal circuit imbalance measure $\kappa_W^*$ (and consequently $\bar{\chi}_A^*$) that grows exponentially with the number of approximation steps.

\section{Preliminaries}

Throughout this paper, we focus on the kernel space $W = \ker(A) \subseteq \R^n$ of the constraint matrix. We rely on the following definitions and results regarding circuits and condition numbers \cite{dadush2024}.

\begin{definition}[Circuits]
For a linear subspace $W \subseteq \R^n$, and a matrix $A$ such that $W = \text{Ker}(A)$, a $\mathbf{circuit}$ is an inclusion-wise minimal dependent set of columns of $A$. Equivalently, a circuit is a set $C \subseteq [n]$ such that $W \cap \R^n_C$ is one-dimensional and no strict subset of $C$ has this property. The set of circuits of $W$ is denoted by $\mathcal{C}_W$. 
\end{definition}

Every circuit $C \in \mathcal{C}_W$ can be associated with a vector $g^C \in W$ such that $\supp(g^C) = C$. This vector is unique up to scalar multiplication.

\begin{definition}[Circuit Ratio and Imbalance Measure]
For a circuit $C \in \mathcal{C}_W$ and $i, j \in C$, we let 
\[\kappa^W_{ij}(C) = \left| \frac{g^C_j}{g^C_i} \right|,\]
For any $i,j \in [n]$, the circuit ratio is defined as
\[
\kappa^W_{ij} = \max \left\{ \kappa^W_{ij}(C) : C \in \mathcal{C}_W, \; i, j \in C \right\}.
\]
where $\kappa_{ij}^{W} = 0$ if there is no circuit supporting $i$ and $j$.
\end{definition}
Note that $\kappa^W_{ij}$ is not necessarily equal to $\kappa^W_{ji}$ for $i \neq j$.

\begin{definition}[Circuit Imbalance Measure]
The $\mathbf{circuit \: imbalance \: measure}$ is \[\kappa_W = \max \{ \kappa^W_{ij} : i, j \in [n] \}.\]
\end{definition}

\begin{definition}[Optimal Circuit Imbalance Measure]
\[\kappa_W^* = \inf_{D \in \mathbf{D}} \kappa_{DW}.\]
\end{definition}

Let $G = ([n], E)$ be the circuit ratio digraph, where $(i,j) \in E$ if $\kappa_{ij}^W > 0$. The arc weights of $G$ are given by $\kappa_{ij}^W$. Dadush et al.\ provide a min-max characterization of the optimal circuit imbalance measure $\kappa_W^*$ that allows it to be lower-bounded by finding cycles in $G$.

\begin{theorem}[Theorem 2.12 in \cite{dadush2024}] \label{thm:cycle}
For a subspace $W \subseteq \R^n$, we have
\[
\kappa_W^* = \max \left\{ \kappa_W(H)^{1/|H|} : H \text{ is a cycle in } G \right\},
\]
where for a cycle $H = (i_1, i_2, \dots, i_k, i_{k+1}=i_1)$, $\kappa_W(H) = \prod_{l=1}^k \kappa^W_{i_l i_{l+1}}$ and $|H| = k$.
\end{theorem}

\subsection{Relating the Optimal Measures}

Theorem 2.8 in Dadush et al.\ \cite{dadush2024} establishes that for any linear subspace $W$,
\[ \bar{\chi}_W \ge \kappa_W. \]

The optimally scaled measures are defined as \[\bar{\chi}_A^* = \inf_{D \in \mathbf{D}} \bar{\chi}_{DW} \: \text{ and } \: \kappa_W^* = \inf_{D \in \mathbf{D}} \kappa_{DW}.\]
Applying the above bound to the rescaled subspace $DW$ yields $\bar{\chi}_{DW} \ge \kappa_{DW}$. Taking the infimum over all $D \in \mathbf{D}$ gives the direct relationship:
\begin{equation} \label{eq:chi_kappa_bound}
\bar{\chi}_A^* \ge \kappa_W^*.
\end{equation}
Thus, demonstrating that $\kappa_W^*$ is exponentially large provides a direct proof that $\bar{\chi}_A^*$ is exponentially large.

\section{The Ben-Tal Nemirovski LP and Standard Equality Form}

The Ben-Tal Nemirovski approximation of the unit disk $D := \{(x,y) : x^2 + y^2 \le 1\} $ using $n$ steps is given by the following linear program:
\begin{align*}
    v_1 &\ge |x|, \quad w_1 \ge |y|, \\
    v_j &= \cos\left(\frac{\pi}{2^j}\right)v_{j-1} + \sin\left(\frac{\pi}{2^j}\right)w_{j-1}, && j=2, \dots, n \\
    w_j &\ge \left| -\sin\left(\frac{\pi}{2^j}\right)v_{j-1} + \cos\left(\frac{\pi}{2^j}\right)w_{j-1} \right|, && j=2, \dots, n \\
    v_n &\le 1, \\
    w_n &\le \tan\left(\frac{\pi}{2^n}\right).
\end{align*}
The purpose of this formulation \cite{bental2001} is to define a polyhedral approximation $\Omega$ by projecting the feasible region from the auxiliary $v$ and $w$ variables onto the original $(x,y)$-space, such that $D$ is bounded as $D \subseteq \Omega \subseteq (1+\epsilon(n))D$, where the error tolerance $\epsilon(n)$ decreases exponentially with the number of approximation steps $n$.

The variables $v, w$ are nonnegative by construction. We split the unconstrained variables into their positive and negative parts: $x = x^+ - x^-$ and $y = y^+ - y^-$. Adding slack variables $c_i, a_{ik}, b_{ij} \ge 0$, we cast this into SEF:
\begin{subequations}
\label{eq:sef}
\begin{align}
    v_n + c_1 &= 1 ,\label{eq:sef_vn} \\
    w_n + c_2 &= \tan\left(\frac{\pi}{2^n}\right) ,\label{eq:sef_wn} \\
    v_1 - (x^+ - x^-) - a_{11} &= 0 ,\label{eq:sef_v1_x1} \\
    v_1 + (x^+ - x^-) - a_{12} &= 0, \\
    w_1 - (y^+ - y^-) - a_{21} &= 0, \\
    w_1 + (y^+ - y^-) - a_{22} &= 0 .
\end{align}
And for $j = 2, \dots, n$:
\begin{align}
    v_j - \cos\left(\frac{\pi}{2^j}\right)v_{j-1} - \sin\left(\frac{\pi}{2^j}\right)w_{j-1} &= 0, \label{eq:sef_vj_recurrence} \\
    w_j + \sin\left(\frac{\pi}{2^j}\right)v_{j-1} - \cos\left(\frac{\pi}{2^j}\right)w_{j-1} - b_{1j} &= 0 ,\label{eq:sef_wj_bound1} \\
    w_j - \sin\left(\frac{\pi}{2^j}\right)v_{j-1} + \cos\left(\frac{\pi}{2^j}\right)w_{j-1} - b_{2j} &= 0 .\label{eq:sef_wj_bound2}
\end{align}
\end{subequations}

We study the kernel $W = \ker(A)$ by setting the right-hand sides of \eqref{eq:sef} to $0$. 

\section{Circuit Analysis}

To construct circuits in $W = \ker(A)$, we restrict the kernel by setting a chosen subset of variables to $0$. We then show that all unassigned variables are uniquely determined up to a single free scalar $t$.

Crucially, we do not need to verify that the remaining slack variables evaluate to strictly non-zero values; we only need to check that they are uniquely determined. If any unassigned variables happen to evaluate to $0$, it simply indicates that the true support of the circuit is a proper subset of the variables. For our purposes, the exact support is immaterial; as long as we explicitly demonstrate that our variables of interest, $v_{n-1}$ and $w_{n-1}$, are strictly non-zero, they are guaranteed to belong to the circuit's support. Furthermore, because the subspace is one-dimensional, their exact circuit ratio is guaranteed by our parametrization.

For the reader's convenience, we note that there are $3n+3$ equations and $4n+8$ variables in the SEF. The circuits we construct set exactly $n+8$ variables to $0$. Thus, we must evaluate exactly $3n$ remaining unassigned variables and verify they are all uniquely determined by $t$.

\subsection{Circuit 1: $\mathcal{C}_1$}

For our first circuit, $\mathcal{C}_1$, we set the following $n+8$ variables to $0$:
\begin{align*}
    b_{1n} &= 0, & b_{2n} &= 0, & c_2 &= 0, & x^- &= 0, & y^- &= 0, \\
    y^+ &= 0, & a_{11} &= 0, & a_{21} &= 0, & a_{22} &= 0, & w_n &= 0,
\end{align*}
and the $n-2$ variables $w_j = 0$ for $j=1, \dots, n-2$.

Setting $x^+ = t$ \textbf{[1 variable]}, we trace the remaining $3n-1$ dependencies in the kernel:
\begin{itemize}
    \item \textbf{Evaluating $v_1$ and $a_{12}$ [2 variables]:} 
    From \eqref{eq:sef_v1_x1}, 
    \[
        v_1 - x^+ - x^- - a_{11} = 0.
    \]
    Because $x^- = a_{11} = 0$, we have $v_1 = x^+ = t$. From 
    \[
        v_1 + x^+ - x^- - a_{12} = 0,
    \]
    we obtain $a_{12} = 2t$.

    \item \textbf{Evaluating $v_j$ up to $n-1$ [$n-2$ variables]:} 
    The recurrence \eqref{eq:sef_vj_recurrence} is 
    \[
        v_j - \cos\left(\frac{\pi}{2^j}\right)v_{j-1} - \sin\left(\frac{\pi}{2^j}\right)w_{j-1} = 0.
    \]
    Since $w_{j-1}=0$ for $j \le n-1$, this simplifies to $v_j = \cos\left(\frac{\pi}{2^j}\right)v_{j-1}$. By induction for $j=2, \dots, n-1$:
    \[
        v_{n-1} = t \prod_{k=2}^{n-1} \cos\left(\frac{\pi}{2^k}\right).
    \]
    Thus $v_{n-1}$ is uniquely determined and strictly non-zero for $t \neq 0$.
    
    \item \textbf{Evaluating $w_{n-1}$ [1 variable]:} 
    Subtracting \eqref{eq:sef_wj_bound2} from \eqref{eq:sef_wj_bound1} at $j=n$ yields 
    \[
        2\sin\left(\frac{\pi}{2^n}\right)v_{n-1} - 2\cos\left(\frac{\pi}{2^n}\right)w_{n-1} = b_{1n} - b_{2n}.
    \]
    Since $b_{1n}=b_{2n}=0$, we have 
    \[
        2\sin\left(\frac{\pi}{2^n}\right)v_{n-1} - 2\cos\left(\frac{\pi}{2^n}\right)w_{n-1} = 0.
    \]
    Rearranging gives:
    \begin{equation} \label{eq:ratio1}
        \frac{v_{n-1}}{w_{n-1}} = \cot\left(\frac{\pi}{2^n}\right).
    \end{equation}
    Thus, $w_{n-1}$ is uniquely determined and non-zero.
    
    \item \textbf{Remaining variables [$2n-2$ variables]:} 
    The remaining unassigned variables are $v_n$ [1 variable], the terminal slack $c_1$ [1 variable], and the recurrence slacks $b_{1j}, b_{2j}$ for $j=2, \dots, n-1$ [$2n-4$ variables]. These depend linearly on the assigned $v$ and $w$ variables, meaning they are uniquely determined by $t$. 
\end{itemize}
Thus all $3n$ unassigned variables are uniquely determined by $t$. So, a subset of these variables defines a valid circuit containing $v_{n-1}$ and $w_{n-1}$.

\subsection{Circuit 2: $\mathcal{C}_2$}

For our second circuit, $\mathcal{C}_2$, we set a different set of $n+8$ variables to $0$:
\begin{align*}
    x^- &= 0, & y^- &= 0, & y^+ &= 0, & a_{11} &= 0, & a_{21} &= 0, \\
    a_{22} &= 0, & c_1 &= 0, & c_2 &= 0, & v_n &= 0, & w_n &= 0,
\end{align*}
and the $n-2$ variables $w_j = 0$ for $j=1, \dots, n-2$.

Letting $x^+ = t$ \textbf{[1 variable]}, we trace the $3n-1$ dependencies:
\begin{itemize}
    \item \textbf{Evaluating $v_j$ up to $n-1$, and $a_{12}$ [$n$ variables]:} 
    This evaluation is identical to $\mathcal{C}_1$, yielding identically parameterized values for $v_1, \dots, v_{n-1}$ [$n-1$ variables] and $a_{12}$ [1 variable].

    \item \textbf{Evaluating $w_{n-1}$ [1 variable]:} 
    Using \eqref{eq:sef_vj_recurrence} at $j=n$ and substituting $v_n=0$, we obtain 
    \[
        0 = \cos\left(\frac{\pi}{2^n}\right)v_{n-1} + \sin\left(\frac{\pi}{2^n}\right)w_{n-1}.
    \]
    Rearranging yields:
    \begin{equation} \label{eq:ratio2}
        \frac{w_{n-1}}{v_{n-1}} = -\cot\left(\frac{\pi}{2^n}\right).
    \end{equation}
    
    \item \textbf{Remaining variables [$2n-2$ variables]:} 
    The remaining unassigned variables are the recurrence slacks $b_{1j}, b_{2j}$ for $j=2, \dots, n$. Unlike $\mathcal{C}_1$, the variables $b_{1n}, b_{2n}$ are not zeroed here. These $2n-2$ variables depend linearly on the assigned variables, uniquely determining them by $t$.
\end{itemize}
Thus all $3n$ unassigned variables are uniquely determined by $t$. So, a subset of these variables defines a valid circuit containing $v_{n-1}$ and $w_{n-1}$.

\section{Condition Number Calculation}

We lower-bound the optimal circuit imbalance measure $\kappa_W^*$ by applying Theorem \ref{thm:cycle}. 

Equation \eqref{eq:ratio1} from circuit $\mathcal{C}_1$ provides a lower bound for the circuit ratio $\kappa^W_{w_{n-1}, v_{n-1}}$ (the ratio of coordinate $v_{n-1}$ to $w_{n-1}$):
\[
    \kappa^W_{w_{n-1}, v_{n-1}} \ge \kappa^W_{w_{n-1}, v_{n-1}}(\mathcal{C}_1) = \left| \frac{v_{n-1}}{w_{n-1}} \right| = \cot\left(\frac{\pi}{2^n}\right).
\]

Similarly, equation \eqref{eq:ratio2} from circuit $\mathcal{C}_2$ bounds the reverse direction:
\[
    \kappa^W_{v_{n-1}, w_{n-1}} \ge \kappa^W_{v_{n-1}, w_{n-1}}(\mathcal{C}_2) = \left| \frac{w_{n-1}}{v_{n-1}} \right| = \left| -\cot\left(\frac{\pi}{2^n}\right) \right| = \cot\left(\frac{\pi}{2^n}\right).
\]

We apply Theorem \ref{thm:cycle} using the cycle $H = (v_{n-1}, w_{n-1}, v_{n-1})$ of length $|H|=2$ in the circuit ratio digraph $G$:
\begin{align*}
    \kappa_W^* &\ge \left( \kappa^W_{w_{n-1}, v_{n-1}} \cdot \kappa^W_{v_{n-1}, w_{n-1}} \right)^{1/2} \\
               &\ge \left( \cot^2\left(\frac{\pi}{2^n}\right) \right)^{1/2} \\
               &= \cot\left(\frac{\pi}{2^n}\right).
\end{align*}

Applying the Taylor series bound $\cot(\theta) \ge  1/\theta - \theta$ for $\theta \in (0, \pi/2]$, we obtain:
\[
    \kappa_W^* \ge \cot\left(\frac{\pi}{2^n}\right) \ge \frac{2^n}{\pi} - \frac{\pi}{2^{n}}.
\]

By \eqref{eq:chi_kappa_bound}, we conclude 
\[
    \bar{\chi}_A^* \ge \cot\left(\frac{\pi}{2^n}\right) \ge \frac{2^n}{\pi} - \frac{\pi}{2^{n}}.
\]

\section{Conclusion}

We showed that the optimal circuit imbalance measure $\kappa_W^*$ for the Ben-Tal Nemirovski approximation of the unit disk grows exponentially with the number of approximation steps $n$. Because $\kappa_W^*$ lower bounds the optimal condition measure $\bar{\chi}_A^*$, this establishes that a natural polyhedral approximation yields exponentially ill-conditioned constraint matrices.

Based on the exponential behavior observed for the Ben-Tal Nemirovski formulation, we suspect that this ill-conditioning is a fundamental property of any such polyhedral approximation of the unit disk, rather than its specific LP construction.

\begin{conjecture}
Let $P = \{ x \mid Ax = b, \, x \ge 0 \}$ be any polyhedron such that $D \subseteq \Pi(P) \subseteq (1+\epsilon)D$, where $D$ is the unit disk in $\R^2$, $\Pi(P)$ is the projection of $P$ onto its first two coordinates, and $\epsilon > 0$. Then $\kappa^*_W \ge \Omega(1/\sqrt{\epsilon})$, where $W = \ker(A)$.
\end{conjecture}

\bibliographystyle{plain}
\bibliography{references}

\end{document}